# Using Random Variables to Predict Experimental Outcomes

By James D. Stein (California State University, Long Beach)

"Prediction is difficult, especially of the future." – Leo Szilard

**Abstract**

We shall show in this paper that there are experiments which are Bernoulli trials with success probability $p \geq 0.5$, and which have the curious feature that it is possible to correctly predict the outcome with probability $> p$.

**Introduction**

How effectively can one predict a Bernoulli trial? It is widely believed that, if one conducts a Bernoulli trial with success probability 0.5, all strategies for predicting outcomes will be successful half the time. If the Bernoulli trial has success probability $p > 0.5$, it is also widely believed that the best one can do is to always predict success, and this will be successful with probability $p$.

We shall show that there is a large class of experiments that appear to the experimenter to have the key characteristics of Bernoulli trials – two outcomes (success and failure) with the same probability of success each time the experiment is conducted – but for which there is a strategy for predicting the outcome that improves on the results of the strategies described in the previous paragraph.

We searched JSTOR for articles using the search terms "prediction" and "Bernoulli trials", ordered by relevance. At the time of the search, there were 743 articles. After going through the first 200 or so, we confess that we wearied of the process, and it is certainly possible that the results in this paper have already been obtained. However, we have talked with several respected mathematicians (mentioned in the Acknowledgements), and they were unaware of this phenomenon.

**Two Contrasting Bernoulli Trial Experiments**

Two people participate in two different experiments. The first person prepares the experiment, and the second person performs the experiment. In the first experiment, the first person takes two fair coins and places them on a table, one to the left of the other. He then flips a fair coin to determine which one of the coins to leave on the table. The second person comes in and flips the coin that remains on the table.

In the second experiment, the first person takes two biased coins. The first coin has a heads probability of 1/3, the second a heads probability of 2/3. He places the two coins on the table, placing the one with a heads probability of 2/3 to the left of the other coin. He then flips a fair coin to determine which one of the coins to leave on the table. The second person comes in and flips the coin that remains on the table.

From the standpoint of the second person, the two experiments are identical. And indeed, if he performs both a sufficient number of times, the flipped coin will land heads approximately half the time in both experiments. However, there is a significant difference between the two experiments – and that difference is the central topic of this paper. As best we can determine at present, there is no known strategy for correctly predicting the outcome of the first experiment more than half the time. However, for the second experiment, which will be analyzed in Section IV of this paper, we will describe a strategy, which we believe traces back to a paper of Blackwell ([1]), which will enable the second person to correctly predict the outcome more than half the time.

Although we have presented the experiments by using an individual who prepared them, this preparation may be provided by hidden variables of which the experimenter is unaware. One of the consequences of the results of this paper is that sometimes the existence of those hidden variables may be brought to light using the strategy we propose.

We shall describe a general framework for such 'predictable' experiments and prove some theorems concerning them.

**Section I – Blackwell's Bet**

Suppose you are shown two envelopes, one white and one brown. There is a different amount of money in each envelope. If you flip a fair coin to determine which envelope to choose, there is a probability of 0.5 that you will choose the envelope with the larger amount.

There is a strategy for improving your probability of guessing which envelope contains the larger sum, which possibly originates in the previously cited Blackwell paper, but is described in a non-technical fashion in ([3]). Open the selected envelope, and choose a continuous random variable from the positive real line that is positive on all open intervals. This random variable will be called a pointer. If the pointer is larger than the amount in the selected envelope, guess that the amount in the other envelope is larger; otherwise guess that the amount in the selected envelope is larger.

The proof that this improves the probability of a successful guess is straightforward. Let L denote the larger amount and S the smaller amount. Let p be the probability that the pointer is less than S, and q the probability that the pointer is greater than L. With probability 0.5, the selected envelope will contain the amount S, and the pointer will direct you to predict that the other envelope has the larger amount with probability 1-p. Similarly, with probability 0.5, the selected envelope will contain the amount L, and the pointer will direct you to predict that the chosen envelope has the larger amount with probability 1-q. The probability of a successful prediction is therefore $0.5(1-p)+0.5(1-q) = 1 – 0.5(p+q) > 0.5$, since $p+q < 1$ as the pointer has positive probability of being in the open interval (S,L).

Notice that something more was needed – something to which the random variable could be compared in order to make a decision. In this example, we used the exact value of the amount of money in the selected envelope to make the comparison that is needed for the prediction. Consider the same experiment, except that instead of placing money in envelopes, we imagine that we are confronted with the problem of determining the larger of two weights. Flip a fair coin to determine which of the two weights to select, let the pointer consists of a third randomly selected weight, and compare the pointer weight and the selected weight by using a balance scale. Use of the balance scale enables us to make the prediction of which weight is heavier without requiring precise knowledge of the values of any of the weights.

Suppose we modify the envelope experiment slightly. Instead of initially placing the two sums of money S and L in the two envelopes in a random fashion, we flip a fair coin; if it lands heads we place the larger sum of money in the brown envelope, if it lands tails we place the larger sum of money in the white envelope. The use of the pointer now enables us to determine which way the coin landed with probability greater than 0.5. This is not a prediction of a coin flip, as the coin has already been flipped.

**Section II – A Ride on the Random Railroad**

We now describe an experiment in which the outcome has a probability of ½, but which can be predicted with a probability greater than ½ before the randomizing device that produces the outcome is used. You find yourself in a railroad station R on an east-west line, having been given (with equal probability) a ticket to the next station either east or west of R, but you have no idea which. You board the train at station R, and one stop later find yourself at either station S1 or S2, with equal probability, in the following diagram.

West <----- S1 ----- R ----- S2 ----- > East

The engineer has a spinner colored red and blue, with the probability r of landing in the red, where 1 > r > ½. The direction that the train now proceeds depends upon the result of the spinner. If the spinner lands in the red, the train will proceed back towards R, if it lands in the blue it will proceed in the opposite direction, going west from S1 or east from S2.

The probability that the train is at station S1 is ½, and from station S1 the probability that it will go east is r. The probability that the train is at station S2 is ½, and from station S2 the probability that it will go east is 1-r. The probability that the train will go east is therefore ½ r + ½ (1-r) = ½.

Think of the track as the real line, and select a random pointer from the real line which is positive on every open interval. Let p be the probability that the pointer is west of station S1, and q the probability that the pointer is east of station S2. If the pointer is east of your present position you will predict that the train will go east, and conversely. The following table enables the computation of the probability of a successful prediction of the train's direction, even before the spinner is used to determine the direction.

| Station | Spinner | Train Direction | Probability Pointer Is Correct | Combined Probability |
|---------|---------|-----------------|--------------------------------|----------------------|
| S1 | red | east | 1 - p | ½ r (1 -p) |
| S1 | blue | west | p | ½ (1-r) p |
| S2 | red | west | 1 – q | ½ r (1 - q) |
| S2 | blue | east | q | ½ (1-r) q |

The probability of the pointer correctly predicting the direction in which the train will proceed is the sum of the entries in the last column. Combining the first and third rows, and the second and fourth rows, this sum is

½ r (2 – (p + q)) + ½ (1-r) (p + q ) = ½ r (2 – 2(p + q)) + ½ (p + q)

$$= r + (p + q) ( ½ - r)$$

$$= r – (r – ½ ) (p + q) > r – (r – ½ ) = ½$$

We shall show later that the use of a pointer is ineffective if all we know about the train is that it is stopped at S1 and will go either east or west on the flip of a fair coin. Considered from the standpoint of the passenger, though, the situation just described, where it is stopped at S1, is indistinguishable from the one where the train is at S1 or S2 with equal probability.  In both cases, the train goes east independently with probability ½ , but in the case where the train is equally likely to be at S1 or S2, the pointer enables one to predict the direction in which the train will proceed with probability > ½.

**"Next Stop – Willoughby"**

   We will again assume you are a passenger in a train on an east-west line.  This time, however, the direction of the train is determined by the flip of a fair coin.  Heads, the train proceeds east to the next station; tails, it proceeds west to the next station.

   It's been a tiring day, and you fall asleep.  You awaken to find yourself in an unknown station, and hear the conductor announce, "Next stop – Willoughby."  You have no idea where Willoughby ([2]) is on the line, and so you assume you are either east or west of it with equal probability.  You use an app on your cell phone to select a random pointer to determine the direction in which the train will proceed.  If the pointer is to your east, you assume the train will go east, and if the pointer is to your west, you assume the train will go west.

   Let p be the probability that the pointer is to the west of the next station to your west, and q the probability that the pointer is to the east of the next station to your east.  An argument virtually identical with the one presented in the section on Blackwell's Bet shows that the probability that you will guess the direction of motion of the train is $1 - 0.5(p + q) > 0.5$.  Notice that successfully guessing the direction of motion of the train is equivalent to successfully guessing the outcome of the coin flip – the coin flip that has already taken place and which determined that the next stop was Willoughby.

However, there was another passenger in the train who obtained the same pointer that you did and made the identical prediction of direction – but this passenger made his prediction BEFORE the coin was flipped to determine the direction. Since he is using the identical pointer as you, and is comparing it to the same location (the station where the train is stopped) as you, the two of you will make identical guesses as to the direction of the train – or the outcome of the coin flip that determines that direction. Therefore, you will have the same probability of successfully guessing the outcome of the flip – a probability we have just shown to be greater than 0.5. But the other passenger is PREDICTING the result of that flip!

**Section III – Extended Bernoulli Trials**

We now introduce a framework which generalizes the type of situation we have been considering, and prove some elementary results.

**Definition -** An extended (two-stage) Bernoulli trial consists of two experiments conducted in sequence. The first experiment has sample space S with outcomes {$x_1$, …. , $x_N$} with probabilities {$p_1$, … , $p_N$} respectively. The second experiment has two outcomes, s and f , and the composite experiment satisfies

$$\sum_{k=1}^{N} p_k P(s|x_k) = p$$

The composite experiment looks like a Bernoulli trial with success probability p. We'll assume p >= 0.5, so that success is the outcome with probability >= 0.5.

Let X be a probability space, and let $E_1, \ldots, E_N$ be subsets of X with probabilities $P(E_1), \ldots, P(E_N)$ respectively. Let v be a random variable defined on X, v is the pointer. If the outcome of the first experiment is $x_k$, we predict success for the compound Bernoulli trial if $v \varepsilon E_k$, and failure if $v \varepsilon X\backslash E_k$. We assume the pointer is independent of the outcome of the first experiment.

The probability for successfully predicting outcomes (abbreviated PSP) is

$$\text{PSP} = \sum_{k=1}^{N} p_k P(s|x_k) P(E_k) + \sum_{k=1}^{N} p_k P(f|x_k) P(X\backslash E_k)$$

We can predict outcomes with results better than chance if PSP > p.

This definition encompasses the critical common aspects of all the examples we have considered – an initial or preparation stage, and a way to incorporate the results of an independent pointer in making a prediction. The two coins in the bag example described in the Introduction, where one coin has a heads probability of 1/3 and the other of 2/3, is an example of an extended Bernoulli trial. The first stage consists of choosing a coin at random from the bag, and the second stage of flipping it.

**Example 1** – We compute PSP for the spinner example of the previous section. The two outcomes $x_1$ and $x_2$ are if the train is at station S1 or S2 respectively, and $p_1 = p_2 = \frac{1}{2}$. Success s will be defined as the train going east. The letters p, q, and r will have the same meaning as in the spinner example. Then

$P(s!x_1) = r \quad P(f|x_1) = 1-r$

$P(s!x_2) = 1-r \quad P(f|x_2) = r$

and

$E_1 = \{ x: x \text{ is east of station S1}\}$     $P(E_1) = 1 - p$

$E_2 = \{ x: x \text{ is west of station S2}\}$     $P(E_2) = 1 - q$

Then

PSP = ½ x r x (1 – p) + ½ x (1-r) x q + ½ x (1 – r) x p + ½ x r x (1 - q) = r – (r – ½ ) (p + q) > ½

as previously computed.

**Notation** – We abbreviate $s_k = P(s|x_k)$ and $y_k = P(E_k)$ for $1 \le k \le N$.

Using this notation, we now obtain a simple criterion for when PSP > p.

$$\text{PSP} = \sum_{k=1}^{N} p_k P(s|x_k)P(E_k) + \sum_{k=1}^{N} p_k P(f|x_k)P(X\setminus E_k)$$

$$= \sum_{k=1}^{N} p_k s_k y_k + \sum_{k=1}^{N} p_k (1 - s_k)(1 - y_k)$$

$$= \sum_{k=1}^{N} p_k (1 + 2 s_k y_k - s_k - y_k)$$

So PSP > p if and only if this last expression is > $\sum_{k=1}^{N} p_k s_k$. So we must have

$$1 + \sum_{k=1}^{N} p_k (2 s_k y_k - s_k - y_k) > \sum_{k=1}^{N} p_k s_k$$

$$1 > \sum_{k=1}^{N} p_k (2 s_k + y_k - 2 s_k y_k) \qquad [1]$$

Notice that when $y_k = 0$, $2 s_k + y_k - 2 s_k y_k = 2 s_k$, and when $y_k = 1$, $2 s_k + y_k - 2 s_k y_k = 1$.

Therefore, if $s_k \ge ½$ for all values of k, it is impossible for inequality [1] to be satisfied.

We shall discuss other situations in which it is impossible for PSP > p in a subsequent theorem. The condition that PSP ≤ p should be interpreted as showing that a particular situation cannot be predicted with better than chance results through the use of a random pointer in the manner we have prescribed. This does not, however, necessarily foreclose on there being other methods of obtaining better than chance results in such situations.

Conversely, being able to show that PSP > p need not necessarily imply that there is a 'useful' scheme for predicting outcomes with results better than chance. In Example 1, there is such a scheme, because all the passenger needs to do while the train is stopped in an unknown station is obtain a random location on the rail line which he can compare with his current position. This could either be done empirically, by observing an event such as a lightning strike and seeing whether it is east or west of his current position. It could also be done by some sort of automated process, such as a cell phone app, which is opaque to the passenger, but simply returns the instruction 'guess east' or 'guess west'.

If one examines the expression for PSP on the left-hand side of the inequality in the line preceding [1], we see that when $y_k = 0$, $2s_k y_k - s_k - y_k = -s_k$, and when $y_k = 1$, $2s_k y_k - s_k - y_k = s_k - 1$. The value of $2s_k y_k - s_k - y_k$ is therefore maximized by choosing $y_k = 0$ if $s_k < ½$, and by choosing $y_k = 1$ if $s_k \geq ½$. These choices therefore maximize PSP.

Notice that this gives us the highest possible PSP by always predicting success when success is more probable, and by always predicting failure when failure was more probable. Of course, this may not always be possible. This highest possible PSP is

$$\sum_{s_k \geq 0.5} p_k\, s_k + \sum_{s_k < 0.5} p_k(1 - s_k) = \sum_{k=1}^{N} p_k s_k + \sum_{s_k < 0.5} p_k(1 - 2s_k) = p + \sum_{s_k < 0.5} p_k(1 - 2s_k)$$

If $s_k < 0.5$, we have $(1-s_k) - s_k = 1 - 2s_k$. We can therefore interpret the last term in the rightmost expression as the maximum premium (in excess of chance) that accrues to the prediction probability if we are able to predict failure for those outcomes from the first stage of the experiment when failure is more likely. For instance, if we have two coins in a bag, one with a heads probability of 0.7 and another with a heads probability of 0.4, and we grab one and flip one, if the two coins appear identical one would obviously predict that the result of the flip would be heads, and this would occur with a probability of 0.55. However, if we are able to tell which coin is which, we would guess "heads" when we picked the coin with heads probability of 0.7, and we would guess "tails" when we picked the coin with heads probability of 0.4. Our successful prediction probability would be 0.5 x 0.7 + 0.5 x 0.6 = 0.65, and this is indeed 0.55 + ½ x (1 − 2 x 0.4) = 0.65, as the previous calculation tells us.

This also agrees with the sentiment outlined in the first paragraph of the Introduction, that in a Bernoulli trial with success probability $\geq$ ½, the best we can do from the standpoint of improving our ability to predict, is to always predict success when success probability is greater than or equal to ½, and always predict failure when success probability is less than ½. We shall prove this shortly for N=1 extended Bernoulli trials (which is the conventional type of Bernoulli trial), but we remark that we have only shown this in conjunction with the use of a random pointer; there may be other ways of improving prediction probability.

Consider this from the standpoint of the two differing experiments with bags containing coins described in the Introduction. If we let outcome 1 denote the coin with heads probability of 2/3, letting $y_1 = 1$ is equivalent to saying that we know we are looking at the coin with heads probability of 2/3, so we always want the pointer to predict success (heads, in this instance). But the intriguing result would be to determine a setup in which PSP > p when we do NOT know which coin we are looking at. We will discuss ways this can be accomplished in Section IV.

In Example 1, however, the passenger has no idea, when the train is stopped in the station, whether outcome 1 or outcome 2 has occurred, i. e. whether he is east or west of Willoughby – but the pointer will enable him to predict whether the train will go east or west with accuracy greater than 50%, and it can do so before the direction of the train has been determined. Of course, this is because of the fact that the train travels east from station S1 more often than it travels east from station S2, and the pointer "detects" this, and thus "knows" in which way the train is likely to move before the train's direction is determined. Here, the pointer is naturally integrated with the problem setup; the pointer is chosen from as a location on the railroad track rather than from some set existing outside the problem.

**Theorem 1 –** If either N=1 or $y_1 = y_2 = \ldots = y_N$, then PSP $\leq$ p.

**Proof:** When N=1, $p_1 = 1$, and there is only $x_1$. PSP = $s_1 y_1 + (1-s_1)(1-y_1) = 1 + 2s_1 y_1 - s_1 - y_1$. If $s = s_1 = \frac{1}{2}$, then PSP = $\frac{1}{2}$. If $s = s_1 > \frac{1}{2}$, then PSP = $1 - s_1 + y_1(2s_1 - 1)$ has its maximum value of $s_1$ when $y_1 = 1$.

If $y_1 = y_2 = \ldots = y_N = y$ and PSP > p, then by [1] we have

$$1 > \sum_{k=1}^{N} p_k(2s_k + y - 2s_k y) = \sum_{k=1}^{N} p_k\big(2s_k + y(1 - 2s_k)\big) = 2p + y - 2py$$

If p = ½, the right side = 1, and 1 > 1, an obvious impossibility. If p > ½, then y(2p-1) > 2p-1, which implies y > 1, which is impossible. QED

The case N=1 in Theorem 1 reinforces our remarks on ordinary Bernoulli trials with success probability greater than or equal to ½. However, the case where $y_1 = y_2 = \ldots = y_N = y$ has an intriguing interpretation. If there were such a value of y, we could simply use that value of y to obtain a pointer which would predict success with greater accuracy than chance no matter which outcome $x_1, \ldots, x_N$ occurred – and we could make that prediction even before stage one of the experiment takes place. This suggests that, in this model, for prediction to be more accurate than chance we need to complete the initial stage, which results in one of the outcomes $x_1, \ldots, x_N$, and then use the pointer (which depends on the outcome). More importantly, to be a true prediction – as in Example 1 – we must not know which outcome has actually occurred.

We conclude this section with an interesting observation that these results raise about the relationship of knowledge to prediction. Numerous examples – hurricane prediction is a notable one, especially in view of the severity of the 2017 hurricane season – rely on the gathering of more knowledge to improve the ability to predict. But extended Bernoulli trials seem to behave differently. Consider the "Next Stop – Willoughby" example. If you know where you are and the next stop depends upon the flip of a fair coin, you are in an N=1 extended Bernoulli trial, and Theorem 1 shows that you cannot use the pointer to improve your prediction capability. However, if you don't know where you are – the N=2 case – you can use the pointer to improve your prediction capability.

What you don't know can help you – at least in some situations.

### Section IV – Extended Bernoulli Trials with N=2

We now look at the case where there are only two initial outcomes $x_1$ and $x_2$. Example 1, "Next Stop – Willoughby", and the bag with two biased coins experiment in the Introduction belong to this important case. Notice that we must have $s_1 < 0.5 < s_2$ to retain the possibility that PSP > p.

**Theorem 2:** If $s_1 < 0.5 < s_2$ and $p = p_1 s_1 + p_2 s_2 \geq 0.5$ and PSP > p, then $y_2 > y_1$,

Proof: Note that we must have $p_2 > 0$. Let $X(y_2) = \sum_{k=1}^{2} p_k (2s_k + y_k - 2s_k y_k)$. Since PSP > p, we must have $X(y_2) < 1$. Notice that $\frac{dX}{dy_2} = p_2(1 - 2s_2) < 0$.

If $y_2 = y_1$, $X(y_2) = 2p + y_1(1-2p) = 2p - y_1(2p-1)$. If $y_1 = 1$, $X(y_2) = 1$, so we must have $y_1 < 1$ to ensure PSP > p. If $y_1 = 0$, $X(y_2) = 2p \geq 1$. So if $y_2 \leq y_1$, since X is a decreasing function of $y_2$, $X(y_2) > 2p - y_1(2p-1) \geq 1$. So $y_2 > y_1$. QED

This result need not hold for N>2, as the following example indicates.

| k | $p_k$ | $s_k$ | $y_k$ | p = 0.56 | PSP = 0.572 |
|---|---|---|---|---|---|
| 1 | 0.2 | 0.3 | 0.1 | | |
| 2 | 0.3 | 0.5 | 0.9 | | |
| 3 | 0.5 | 0.7 | 0.7 | | |

We can actually do a little better. If p > 0.5 in Theorem 2, then $y_1 < y_2$ is not sufficient to ensure that PSP > p. If $y_1 = y_2 = 0$, then the right side of [1] is $2 p_1 s_1 + 2 p_2 s_2 = 2p > 1$, so by continuity we can find $0 < y_1 < y_2$ with the right side of [1] greater than 1. If p = ½ and $y_1 = y_2 = 0$ or $y_1 = y_2 = 1$, then the right side of [1] equals 1, so $y_1 = y_2$ implies that the right side of [1] equals 1. But, as above, $\frac{dX}{dy_2} = p_2(1 - 2s_2) < 0$, so increasing $y_2$ will ensure that $y_1 < y_2$ implies the right side of [1] is less than 1.

To summarize, if p>0.5, $y_1 < y_2$ will not guarantee that PSP>p, but if p=0.5, it will.

Notice what happens if we model a two-outcome extended Bernoulli trial by using the unit interval with the uniform distribution. Outcome $x_1$ will be a position on the line such that the selection of a random number greater than that position will be considered a success; similarly for $x_2$. Since $s_1 < s_2$, with this interpretation the location of $x_2$ is smaller than the location of $x_1$. If the unit interval is the model for a physical parameter such as length or weight, the outcome with the higher probability for success must occur prior to (lower value of parameter) the outcome with the lower probability for success.

**Physical Models Ensuring PSP > p for Two Biased Coins in a Bag**

From a probabilistic or statistical standpoint, there is no distinguishable difference between the outcomes of sequences of flips in the two experiments described in the Introduction. It is easier to see this if, instead of biased coins, we use jars filled with black and white marbles.

Suppose we have two jars of marbles. The first contains 100 white and 50 black marbles, and the second contains 50 white and 100 black marbles. We claim that choosing a jar and then choosing (with replacement) a marble from that jar will generate the same sequences of black and white marbles that choosing (with replacement) a marble from a jar containing 150 white and 150 black marbles would.

Put all 300 marbles from the two jars in a box with two sides separated by a partition. On one side are 50 black and 100 white, on the other side are 50 white and 100 black. In reaching in and drawing a marble from the whole box, you are doing the same thing as reaching in to one side or the other and picking a marble from that side. And vice versa, in selecting a side and then drawing a marble from that side, you could be picking any random marble from the whole box. So, simply from the standpoint of either probability or statistics, the outcomes of sequences of flips from either the two jars with 150 marbles or the single jar with 300 marbles would be indistinguishable.

However, it is possible to set up the experiment with two jars in such a way that we can ensure that PSP > p. We describe the experiment using two biased coins as in the Introduction, but jars of marbles or other physical experiments with the appropriate probabilities will work equally well. We assume that Coin 1 has a heads probability of $s_1 < ½$, and Coin 2 has a heads probability of $s_2 > ½$, where $s_1 + s_2 = 1$. There are two people involved, the preparer who sets up the experiment, and the experimenter who performs the experiment. In these examples, the experimenter always flips a coin and predicts whether it will land heads or tails. It "feels" like an ordinary Bernoulli trial to him.

**Model 1 (pointer uses position)** – The preparer places two coins on a table, making sure that Coin 2 is to the left of Coin 1. Flip a fair coin to determine which coin to leave on the table. The experimenter then chooses a random location on the table (this can also be done prior to the setting up of the experiment). If the random location is to the right of the remaining coin, the experimenter predicts that the flip will land heads, otherwise he predicts that the flip will land tails. The experimenter then flips the remaining coin.

**Model 2 (pointer uses mass)** – The preparer places two coins on a table, making sure that Coin 2 is the lighter of the two. He then flips a fair coin to determine which coin to leave on the table. The experimenter then chooses a random weight to compare with the remaining coin; as before this can be done prior to the setting up of the experiment. If the random weight is heavier than the remaining coin, the experimenter predicts that the flip will land heads, otherwise he predicts that it will land tails. The experimenter then flips the remaining coin.

**Model 3 (pointer uses time)** – The preparer places a safe with a time lock on a table, and chooses two times at random. The experimenter chooses a third random time at which he will walk into the room. The preparer then flips a fair coin to determine which of the two random times to choose. If the chosen time is the earlier time, place Coin 2 in the safe, and set the time lock to the earlier time. If the chosen time is the later time, place Coin 1 in the safe, and set the time lock to the later time. The experimenter walks into the room at the time he randomly determined. If the time lock allows the safe to be opened, predict that the flip will land heads, otherwise predict the coin will land tails. The experimenter then flips the coin when the time lock allows him to do so.

In each of Models 1 to 3, $y_k$ is the probability that the pointer will be larger than the parameter assigned to Coin k (distance from left of table in Model 1, mass in Model 2, time of placement on top of table in Model 3), and in all three cases, $y_2 > y_1$. Clearly $p = (s_1 + s_2)/2 = ½$.

Furthermore, inequality [1] is satisfied because if $y_1 = y_2$, then

$0.5(2 s_1 + y_1 − 2 s_1 y_1 + 2 s_2 + y_2 − 2 s_2 y_2) = 0.5(2 s_1 + y_1 − 2 s_1 y_1 + 2 s_2 + y_1 − 2 s_2 y_1)$

$= 0.5(2 + y_1(2 − 2 s_1 − 2 s_2)) = 1$

Again, letting $X(y_2) = 0.5(2 s_1 + y_1 − 2 s_1 y_1 + 2 s_2 + y_2 − 2 s_2 y_2)$, we see that $\frac{dX}{dy_2} = 0.5 (1 − 2s_2) < 0$, and so if $y_2 > y_1$, then $X(y_2) < 1$, satisfying the inequality [1].

Recalling the discussion following Theorem 2, it is the flip of the fair coin that ensures that $p_1 = p_2$, and together with the requirement that $s_1 + s_2 = 1$ yield the result that PSP > p. The motivation for these restrictions is to assure that PSP > p as long as we choose $y_1 < y_2$. Of course, if one were willing to place constraints on $y_1$ and $y_2$, one can sometimes guarantee that PSP > p in cases where $p_1 \neq p_2$ or $s_1 + s_2 \neq 1$.

It must be admitted that there is something very definitely artificial about Models 1 through 3, when contrasted with Example 1. In Models 1 through 3, the experiment and the pointer are forcibly linked through restrictions on both the experiment and identifications, whereas in Example 1 the experiment and the pointer blend seamlessly with each other along the rail line.

**Section IV – Concluding Comments**

Assuming the results of this paper are validated (and we have written a number of simulations which appear to substantiate them), the field is wide open for expanding and potentially using these results.

One particular area of interest would be physics, especially quantum mechanics. There are some instances in physics that appear to be extended Bernoulli trials, as probabilistic descriptions of phenomena are well known. The author is not qualified to comment in this area, but would welcome input from physicists.

A second use would be to disclose "hidden variable" type situations as described earlier in the Introduction.

Finally, lifting the veil that clouds prediction – even to the small extent that has been demonstrated in this paper – raises interesting mathematical and philosophical questions.

**Acknowledgements**

This paper has benefited from discussions with Bruce Rothschild, Phil Everson, and Art Benjamin, and the author is grateful for those discussions. But this paper would not even have come into existence without the author having had numerous fruitful, enjoyable, and intriguing conversations with Len Wapner, who first made the author aware of Blackwell's Bet, and whose insight and ability to ask simple but profound questions have motivated much of this paper.

James D. Stein

Department of Mathematics and Statistics

California State University, Long Beach

Long Beach, CA

james.stein@csulb.edu